# A review of the variational methods for solving DC circuits

N R Sree Harsha

School of Nuclear Engineering,
Purdue University, USA.
*Email: snaropan@purdue.edu*

**Abstract**

Direct current (DC) circuits are usually taught in upper-level physics curricula, and Kirchhoff's laws are stated and used to solve the steady-state currents. However, students are not often introduced to alternative techniques, such as variational principles, for solving circuits. Many authors have tried to derive the steady-state distribution of currents in circuits from variational principles, and the initial attempts were carried out by great physicists, such as Kirchhoff, Maxwell and Feynman. In this article, we shall review such variational principles in physics and illustrate how they can be used to solve DC circuits. We will also explore how they are related, in a fundamental way, to the entropy production principles of irreversible thermodynamics. We believe that the introduction of variational alternatives for solving circuits can be a good opportunity for the instructor to present the techniques of the calculus of variations and irreversible thermodynamics to students.

## 1. Introduction

The calculus of variations, first formulated by Wilhelm Leibniz, represents a powerful alternative to the differential approach developed by Newton in solving the dynamics of classical mechanical systems [1]. While the Newtonian approach is based on solving second order differential equations and involves vector quantities like velocity, displacement, force etc., the variational method to classical mechanics underlies identifying a path that makes an integral quantity stationary. The calculus of variations can, in a broad sense, be defined as the mathematical study of finding maximum, minimum and critical points of a function $f : M \to \mathbb{R}$, where $\mathbb{R} = \{\text{Real numbers}\}$. In the general case, *M* can be a set of numbers, functions, paths, surfaces etc. If *f(x)* is a function of a single variable *x*, the critical points of the function *f(x)* are found by setting *df(x)/dx = 0* and these critical points are minimum, maximum or inflection points depending on whether $d^2f/dx^2$ is positive, negative or zero respectively. These results also work for functions with more than one variable such as $f(x_1,x_2,x_3...x_n)$ and the respective critical points corresponding to a variable $x_i$ can be found by setting $df/dx_i = 0$.

One of the first problems that formed the foundations for the study of the calculus of variations is finding the curve that has the shortest descent time, called the *brachistochrone* problem. This problem was first published by Johann Bernoulli in June 1696 [2]. Johann Bernoulli's own solution utilized Fermat's optical principle of least time and was published in May 1697 [3]. The explanation of the problem and solution in modern terms is presented in the excellent papers of Herman Erlichson [4] and Henk Broer [5]. The calculus of variations was later developed rigorously by Lagrange and Euler into a mathematical discipline of finding solutions to general extreme value problems. It was later realized, by many great mathematical physicists such as Hamilton, D'Alembert, Paul Dirac, Feynman etc., that many general laws of classical and quantum mechanics can be formulated into a compact form, that is now called the principle of least (or, more precisely, stationary) action. We shall now state the principle and show how it can be used to arrive at the Newton's equations of motion for a simple classical system.

The principle of least action states that, with end points fixed between sufficiently short time intervals, the true trajectory employed by the system is the one that minimizes a quantity, called action $A$. For larger time intervals, the true trajectory may be a saddle point of the functional [6]. The action is usually written out as

$$A = \int_{t_0}^{t_1} L(q_i, \dot{q}_i, t) dt \qquad (1)$$

The functional $L$ in equation (1) is a function whose arguments are themselves functions and $q_i$ represents the generalized coordinate of the $i^{th}$ particle, $t$ represents the time coordinate with $\dot{q}_i = dq_i/dt$. The action in equation (1) is stationary when $\delta A = 0$. To find the critical points of the functional $A$, the Lagrange method of multipliers are often used. Since this technique will also be used later in the article, a brief overview of the method is now presented. The conditional extrema of a function $f(x)$ of many variables $x_1, x_2, x_3 \ldots x_n$, subject to $m$ equations of constraints $\varphi_1(x_1, x_2 \ldots x_n) = 0$, $\varphi_2(x_1, x_2 \ldots x_n) = 0 \ldots \varphi_m(x_1, x_2 \ldots x_n) = 0$, is equal to the absolute extrema of the function $F(x)$ such that

$$F(x) = f(x) + \sum_{i=1}^{m} \lambda_i \varphi_i(x_1, x_2 \ldots x_n) \qquad (2)$$

Here, $\lambda_i$ are called the Lagrange multipliers and $F(x)$ is called the Lagrange function of $f(x)$. The absolute extrema of the Lagrange function $F(x)$ can be found for each variable $x_i$ by setting $\frac{\partial F(x)}{\partial x_i} = 0$. Hence, the method of finding conditional extrema of a function $f(x)$ has been reduced to finding absolute extrema of its Lagrange function $F(x)$. Using this technique of finding the conditional extremum, it can be shown [7] that $\delta A = 0$ along a curve $\gamma: q = q(t)$ passing through the end points $q(t_0) = q_0$ and $q(t_1) = q_1$ if, and only if,

$$\frac{d}{dt}\frac{\partial L}{\partial \dot{q}_i} - \frac{\partial L}{\partial q_i} = 0 \qquad (3)$$

The condition shown in equation (3) is called the Euler-Lagrange condition and finding the path along which the action is stationary is called the principle of least (or, more precisely, stationary) action. We shall now demonstrate its application through a simple example.

The Lagrangian for a single particle, with kinetic energy $T$ and a potential energy $V$, is given by $L = T - V$. The nonrelativistic kinetic energy for the particle is given by

$$T = \frac{1}{2}m\dot{x}^2 \qquad (4)$$

The Lagrangian $L$ for this special case is then given by

$$L = \frac{1}{2}m\dot{x}^2 - V \qquad (5)$$

Next, applying the Euler–Lagrange condition shown in equation (3), for Lagrangian defined in equation (5), and using the familiar definition of the force on the particle as $F = -dV/dx$, gives us the familiar Newton's formula $F = m\ddot{x}$. The principle of least action is a powerful tool to represent various theories in physics and different Lagrangian functionals can be formulated that neatly encapsulate Einstein's theory of general relativity, Maxwell's electromagnetic theory, the standard model of elementary particles etc. [8].

The calculus of variations can also be used to solve other problems in classical mechanics such as the isoperimetric, catenary problems [9] and the problem of turning quickly in the least amount of time [10]. The Fermat's principle of least time in Optics, formulated in the 17th century, also represented a guiding principle in formulating the physical laws using the variational calculus. In the recent years it has also been suggested that Fermat's principle can be used to explain the extremely low velocities of light in Bose-Einstein condensates [11]. This can then be used to create an analog of a black hole to study the behavior of light paths around the event horizon. A considerable success was achieved in applying variational principles in quantum mechanics to solve problems such as calculating the energy levels of the Helium atom [12], in the BCS theory of superconductivity [13] and Feynman's theory of superfluid Helium [14]. The variational approach to classical thermodynamics was also formulated by H. A. Buchdahl, who showed that for quasi-static transitions the Second law of thermodynamics can be formulated as a variational principle [15]. The equations of motion for the damped harmonic oscillator was also formulated using variational principles in classical and quantum mechanics [16].

In equilibrium thermodynamics, however, it is well known that a function called *entropy* will be maximized as the system reaches the equilibrium state[1]. Ehrenfests were the first to ask if there exists an analogous function that achieves an extremum value when a system achieves a stationary non-equilibrium state [17]. Many attempts were made in the last century to discover such a function. We will, in this paper, review such methods done in the past to determine the solutions of the Direct Current (DC) circuits. The DC circuits are systems of voltage sources and resistors introduced early in the undergraduate physics and electrical engineering courses. In this review paper, we shall treat the DC circuits as the systems obeying the laws of non-equilibrium thermodynamics. Consider, as an example to demonstrate an advantage of such a treatment, Ohm's law. It is well known that the electrons in the wire in a simple DC circuit are governed by the electric fields present inside the wire. These fields inside the wire are created due to the surface charges present on the wire and charges present on the terminals of the battery [18]. However, students are taught in electrostatics that electric fields cannot exist inside a conducting wire. Hence, the question that could then arise to the inquisitive mind is this: how can an electric field exist inside a current carrying wire, which is part of a larger DC circuit? The answer is that a DC circuit is in a non-equilibrium steady state, whereas the argument from electrostatics is only valid for steady equilibrium states.

---

[1] It should be noted that entropy maximization in equilibrium states is not a variational principle but is an extremum condition.

In this paper, we will predominantly be concerned with the possibility of solving DC circuits through the variational methods developed in the non-equilibrium thermodynamics. Kirchhoff's laws, however, are often used to solve for the currents in different branches of a linear, planar, DC electric circuits [19]. As a simplifying assumption, we shall treat the network components as lumped network parameters in which the propagation delay of electromagnetic signals is ignored. The Kirchhoff's laws are divided into the voltage law (KVL) and the current law (KCL). Alternative methods for finding the currents in different branches of a resistive DC electric circuit have been proposed by many authors. In sections 2 and 3, we shall consider such attempts and in sections 4 and 5 we shall see how entropy production principles lie at the heart of these variational methods. The limitations of using these variational methods are presented in section 6 and the pedagogical advantages are outlined in section 7. In what follows, we will consider all the voltage sources to be ideal and the resistors to be linear, obeying the Ohm's law Since we are more interested in application of variational principles in circuits, we shall only solve them in the case of simple circuits, but extending them to more general cases is trivial.

## 2. Kirchhoff and Maxwell

Kirchhoff's current law states that currents entering a node are equal to currents leaving a node in a circuit. This also means that there can be no accumulation of charges at any point in the steady state operation of a DC circuit. We shall now consider a three-dimensional circuit in which the conductivity $\sigma$ is a constant throughout the region of interest. The generalization of Kirchhoff's current law in the three-dimensional case is that the divergence of the current density $\vec{J}$ is equal to zero. Thus, we have

$$\nabla \cdot \vec{J} = 0 \qquad (6)$$

If the electric field is represented by $\vec{E}$, we can write the current density as the product of conductivity and electric field at a point in space. Hence, we have

$$\vec{J} = \sigma \vec{E} \qquad (7)$$

The electric field can be written as the negative gradient of the scalar potential function $\phi$ i.e.

$$\vec{E} = -\nabla \phi \qquad (8)$$

Hence, from equations (6), (7) and (8), we get

$$\nabla \cdot (\sigma \nabla \phi) = 0 \qquad (9)$$

If we assume the conductivity to be constant throughout the considered space, we get

$$\nabla^2 \phi = 0 \qquad (10)$$

Hence the steady state in a DC circuit is achieved when equation (10) is satisfied at all the points in the circuit. Next, the rate of production of heat $P$ in a volume $V$ surrounded by a boundary $S$, is given by

$$P = \int_V \sigma (\nabla \phi)^2 \, dV \tag{11}$$

Applying the divergence theorem and making use of equation (10), we get

$$P = \int_V \nabla \cdot (\sigma \phi \nabla \phi) \, dV = \int_S (\sigma \phi \nabla \phi) \cdot d\vec{S} \tag{12}$$

If the conductivity is a constant throughout the volume $V$, the variation of the rate of heat production $P$ in $V$ is given by

$$\delta P = \sigma \int_S (\delta \phi \nabla \phi) \cdot d\vec{S} + \sigma \int_S (\phi \nabla (\delta \phi)) \cdot d\vec{S} \tag{13}$$

It can be seen from equation (13) that if variations in flux, $\delta\phi$, vanish on the boundary of $V$ then $P$ is stationary, i.e. $\delta P = 0$. Hence, electric currents are distributed in the region, with an applied voltage on its boundary, so that the rate of production of heat is extremal for the stationary state. Kirchhoff noted this in his paper published in 1848 [20]. This is, as E. T. Jaynes stated, probably the first example in determining a variational principle in steady-state non-equilibrium thermodynamics [21]. Maxwell applied this principle in the context of electric circuits represented by lumped parameters, which is demonstrated in the following example [22].

Consider a simple DC circuit shown in Figure 1. The ideal current source $I$ is connected to a parallel combination of two linear ideal resistors $R_1$ and $R_2$. We will arrive at an equation for individual currents $i_1$ and $i_2$, flowing through the resistors $R_1$ and $R_2$ respectively, by applying KCL at node **A** and by minimizing the total power $P$ dissipated in the two resistors. We shall imagine a closed boundary $S$ surrounding the parallel combination of resistors in the circuit. Since, in the steady nonequilibrium operation of the circuit, the potential differences across the components are fixed, we must then have that the variation of potential on the surface of the boundary $S$ should vanish. As demonstrated in the three-dimensional case, this implies that the power dissipated in the region surrounded by $S$ must be stationary.

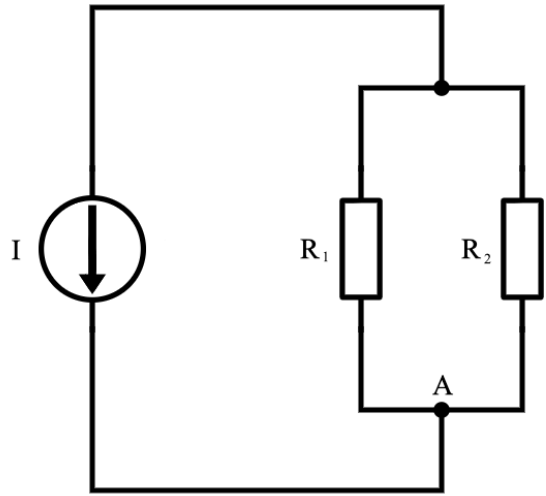

Figure 1. A simple circuit to demonstrate Maxwell's idea.

Applying KCL at node A, we get $i_1+i_2=I$. The total power dissipated in the resistors is

$$P = i_1^2 R_1 + (I - i_1)^2 R_2 \tag{14}$$

Since, the only variable in equation (14) is current $i_1$ through resistor $R_1$, minimizing the power dissipated with respect to the current $i_1$, we have

$$\frac{dP}{di_1} = 2i_1 R_1 - 2(I - i_1) R_2 = 0 \tag{15}$$

Solving equation (15), we get the correct distribution of currents in the circuit as

$$i_1 = \frac{IR_2}{R_1 + R_2} \text{ and } i_2 = I - i_1 = \frac{IR_1}{R_1 + R_2} \tag{16}$$

In order to prove that the Joule power dissipated in the resistors is a minimum, we use equation (15) again to get

$$\frac{d^2 P}{di_1^2} = 2(R_1 + R_2) > 0 \tag{17}$$

It should be noted that we have solved for the currents in the circuit, shown in figure 1, without the application of KVL. It should also be noted that this method of solving for unknown currents is applicable to any arbitrary closed DC circuit only when current sources are present. An analogous method of finding steady-state currents in the circuit was obtained by A. A. P. Gibson and B. M. Dillon when only voltage sources are present in the circuit [23]. In order to illustrate Gibson and Dillon's method, we consider a simple circuit shown in figure 2.

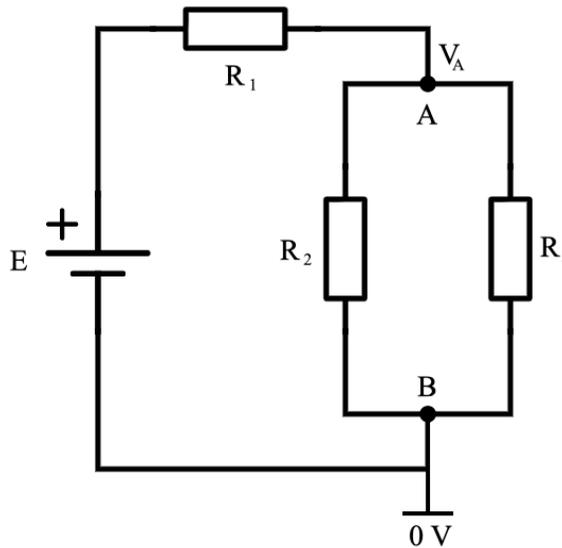

Figure 2. A simple circuit to demonstrate Gibson and Dillon's idea.

Let us assume that current $i$ flows through the resistor $R_1$ and the ideal voltage source $E$, the voltage at node A is $V_A$ and the voltage at the node B is 0 Volts (reference voltage[2]). The power $P$ dissipated in all the resistors can then be written as

$$P = \frac{(E-V_A)^2}{R_1} + \frac{V_A^2}{R_2} + \frac{V_A^2}{R_3} \qquad (18)$$

We now vary the node voltage $V_A$ and find the solution at which $P$ is stationary. Analytically, this can be done by setting $dP/dV_A=0$. Hence, we have

$$\frac{dP}{dV_A} = -\frac{2(E-V_A)}{R_1} + \frac{2V_A}{R_2} + \frac{2V_A}{R_3} = 0 \qquad (19)$$

Solving the equation (19) for $V_A$ gives us that

$$V_A = \frac{ER_2R_3}{\sum R_1 R_2} \qquad (20)$$

The current $i$ is then given by Ohm's law as

$$i = \frac{E-V_A}{R_1} = \frac{E(R_2+R_3)}{\sum R_1 R_2} \qquad (21)$$

This is the correct distribution of currents in the steady-state operation of the circuit. In the next section, we shall describe a variational method for solving circuits that contain arbitrary combination of the current and voltage sources.

## 3. J. J. Thompson and J. Jeans

The idea of a quantity, whose extremum determines the steady state distribution of currents in an arbitrary combination of resistors and voltage sources, was first presented by J. J. Thompson when he stated that "[w]e can prove in a similar way that when there are electromotive forces in the different branches the currents adjust themselves so that $\sum RC^2 - 2\sum EC$ is a minimum, where $E$ is the electromotive force in the branch when the current is $C$" [24]. Shortly thereafter, J. Jeans also noted that "[w]hen a system of steady currents flows through a network of conductors of resistances $R_1$, $R_2$, ..., containing batteries of electromotive forces $E_1$, $E_2$, ..., the currents $x_1$, $x_2$, ... are distributed in such a way that the function $\sum Rx^2 - 2\sum Ex$ is a minimum, subject to the conditions imposed by Kirchhoff's first [current] law; and conversely" [25]. We illustrate their ideas through the circuit shown in figure 3.

---

[2] The reference voltage need not necessarily be 0 Volts. A non-zero constant reference voltage, however, does not change our analysis, since we will only be interested in $dP/dV_A$.

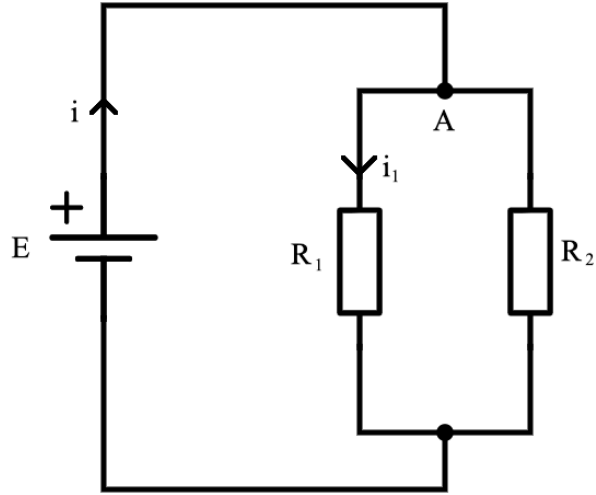

Figure 3. A simple circuit to demonstrate J. J. Thompson's idea.

Let the current from the ideal voltage source $E$ by $i$ and the current through the ideal resistor $R_1$ by $i_1$. The application of KCL at node A requires that the current through the ideal resistor $R_2$ be $i - i_1$. Hence, the quantity to be minimized is

$$S_{JJ} = \sum Rx^2 - 2\sum Ex = i_1^2 R_1 + (i - i_1)^2 R_2 - 2Ei \qquad (22)$$

In order to minimize $S_{JJ}$ with respect to its independent parameters $i$ and $i_1$, we must have that

$$\frac{\partial S_{JJ}}{\partial i} = \frac{\partial S_{JJ}}{\partial i_1} = 0 \qquad (23)$$

Applying the conditions shown in equation (23), we have

$$i = \frac{E(R_1 + R_2)}{R_1 R_2} \qquad (24)$$

For the individual currents $i_1$ and $i - i_2$ flowing through resistors $R_1$ and $R_2$ respectively, we have

$$i_1 = \frac{iR_2}{R_1 + R_2} \qquad (25)$$

$$i - i_1 = \frac{iR_1}{R_1 + R_2} \qquad (26)$$

Hence, we have solved the circuit shown in figure 3 without the application of KVL. It should be noted that J. J. Thompson's method is also applicable when a combination of current sources and voltages sources are present in the circuit. When the current sources are also present along with the voltage sources, the method to calculate the distribution of steady state currents was given explicitly by D. A. Van Baak [26]. We shall only state Van Baak's theorem here and, for a detailed proof of his theorem with pedagogical advantages of teaching it, the reader is referred to his excellent article (ref. 26).

***Van Baak's theorem***: In an arbitrary combination of voltage sources, current sources and ohmic resistors, the true distribution of currents is the one which extremizes the quantity

$$S_{Van} = P_d - 2P_g \tag{27}$$

Here $P_d$ represents the rate of ohmic dissipation and $P_g$ represents the rate at which all voltage sources do work on the currents. It should be noted that the rate at which current sources do work is not taken into account when formulating $S_{Van}$. We shall now see how these variational principles emerge from the energy balance equation from Tellegen's theorem [27]. Consider an electric circuit with *n* branches with instantaneous voltages of $V_1$, $V_2$, $V_3$ ... $V_n$ across them and an instantaneous current of $i_1$, $i_2$, $i_3$ ... $i_n$ flowing through them. Then Tellegen's theorem states that the directed sum of voltages and currents over all the branches is zero i.e.

$$\sum_{i=1}^{n} V_i i_i = 0 \tag{28}$$

Alper Ercan, applying Tellegen's theorem, recently showed that in an electric circuit, "energy conservation, KCL and KVL are strongly interrelated: any two together imply the other" [28]. As an illustrative example, consider the circuit shown in figure 4.

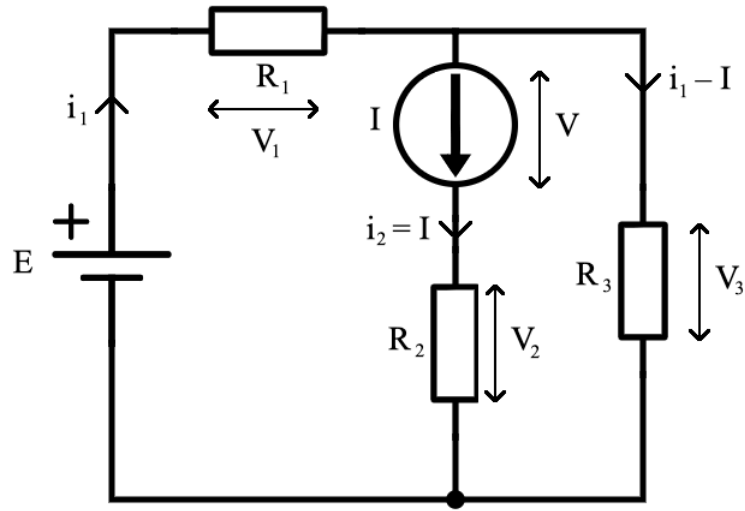

Figure 4. A simple circuit to demonstrate the equivalence of Alper Ercan's and J. J. Thompson's methods.

If the voltage across the current source *I* is *V*, the voltages across the resistors $R_1$, $R_2$ and $R_3$ are $V_1$, $V_2$ and $V_3$, respectively, the energy balance equation reads

$$-Ei_1 + V_1 i_1 + V_2 i_2 + VI + (i_1 - I)V_3 = 0 \tag{29}$$

Next, a variational quantity $S_{AE}$ that will give us the correct distribution of steady-state distribution of currents in the circuit can be constructed as

$$S_{AE} = -\int_0^{i_1} E\, di_1 + \int_0^{i_1} V_1 di_1 + \int_0^{(i_1-I)} V_3\, d(i_1-I) \qquad (30)$$

Using Ohm's law at resistors $R_1$ and $R_3$, we get

$$S_{AE} = -Ei_1 + \frac{i_1^2 R_1}{2} + \frac{(i_1-I)^2 R_3}{2} \qquad (31)$$

Since $I^2 R_2$ is a constant quantity, we can multiply $S_{AE}$ by 2 and add $I^2 R_2$ to arrive at the variational quantity discovered by Van Baak for DC circuits consisting of current and voltage sources i.e., from equations (31) and (27), we have $2S_{AE} + I^2 R_2 = S_{Van}$. The stationary solution can then be found by setting $dS_{AE}/di_1 = dS_{Van}/di_1 = 0$ and this gives us the correct distribution of current as $i_1 = (E+IR_3)/(R_1+R_3)$.

### 4. Minimum Entropy Production principle

Feynman, in his book [29], stated that "[a]s an example, if currents are made to go through a piece of material obeying Ohm's law, the currents distribute themselves inside the piece so that the rate at which heat is generated is as little as possible. […] The new distribution [of currents in the wire] can be found from the principle that it is the distribution for a given current for which the entropy developed per second by collisions is as small as possible." Taking this as a clue, José-Philippe Pérez reasoned that the minimum of electrical power in a DC circuit must somehow be related to the minimum of entropy production [30]. In formulating a variational principle for non-equilibrium or irreversible phenomena, we make use of Onsager's reciprocity theorem [31, 32]. A first step in formulating a variational principle for irreversible thermodynamics was taken by Prigogine, when he formulated the principle of Minimum Entropy Production (MinEP) for thermodynamic systems close to equilibrium [33]. We shall now review some concepts required to formulate a variational principle for DC circuits.

In order to develop such a variational principle, we need a working definition of entropy in electric circuits consisting of resistors. As an example, consider an isolated DC electric circuit consisting of only ideal voltage sources, represented by $E$, and Ohmic resistors, represented by $R$, in an environment that is at a constant ground potential and at held at a constant temperature $T_a$. The entropy is usually defined for the steady-state of a system in equilibrium. In order to define entropy for a non-equilibrium system, like a DC circuit, we make use of forces and fluxes to describe the state of the system [34]. A force is a generalized function of the intensive parameters of the non-equilibrium system that "drives" the system toward equilibrium. Such generalized forces, represented as $F_k$, are also called as affinities. We define a current density $\vec{J}_k$ of an extensive parameter $X_k$ to be the response of the system subject to the forces. Next, the Onsager's linear relationship states that the thermodynamic forces and fluxes are linearly related by a form given by [35]

$$J_i = \sum_k L_{ik} X_k \text{ with } L_{ik} = L_{ki} \tag{32}$$

The entropy production rate per unit volume, $\dot{s}$, is given by

$$\dot{s} = \sum_k \nabla F_k \cdot \vec{J}_k \tag{33}$$

The electric thermodynamic force is given by $F_e = \nabla\left(-\dfrac{\phi}{T}\right)$. For a circuit operating at a constant temperature of $T_a$, the rate of production of entropy per unit volume then is given from equations (33) and (8) as

$$\dot{s} = \dfrac{-\vec{J}_e \cdot \nabla \phi}{T_a} = \dfrac{\vec{J}_e \cdot \vec{E}}{T_a} \tag{34}$$

The total entropy produced in a resistive element of the circuit is then given by

$$\delta S = \dfrac{dt}{T_a} \int_V \dfrac{J^2}{\sigma} dV = \dfrac{RI^2 dt}{T_a} = \dfrac{P_d dt}{T_a} \tag{35}$$

Here, the current that flows through the resistors in the circuit is given by $I$. Similarly, the entropy produced in an electric generator is given by $\delta S_g = -\dfrac{EI dt}{T_a}$. As José-Philippe Pérez proved in his article [30], the total rate of the entropy production in a DC circuit with a generator is given by $\dfrac{dS}{dt} = \dfrac{(P_d - 2P_g)}{T_a}$. Thus minimizing rate of entropy production in the circuit reduces to minimizing the variational quantity $P_d - 2P_g$ as described in section 3.

    The MinEP production states that the entropy production in a system, subject to constant irreversible forces $X_i$ and Onsager's linear relationship, is a minimum [36]. We shall illustrate the use of the MinEP principle in solving a simple DC circuit shown in figure 5. In order to apply the MinEP principle, we must consider a subsystem and define the irreversible forces acting on it. For our purposes, we shall consider the parallel combination of resistors $R_2$ and $R_3$ as a subsystem and we shall denote it by $\Omega$. The KCL is assumed to be applicable at every point in the circuit and hence, we assume that currents $i_1$ and $i - i_1$ flow through the resistors $R_2$ and $R_3$. Since the thermodynamic forces in an electric circuits are defined by $F_e = \nabla\left(-\dfrac{\phi}{T}\right)$, where $\phi$ is the potential across voltage source in the circuit held at a room temperature $T$, we must also have that the electric field, given by $\nabla(-\phi)$, in the voltage source is a constant. From the generalized Ohm's law, we must then also have that the current density, given by $\vec{J}_e = \sigma(-\nabla\phi)$, must also be a constant (here $\sigma$ is the local conductivity of the material in the voltage source). Hence, for the applicability of MinEP in the subsystem $\Omega$, we must impose a condition that the current supplied by the voltage

source is a constant. In our particular example, we must then have that $i$ is a constant, while we vary rate of entropy production in $\Omega$ with respect to $i_1$.

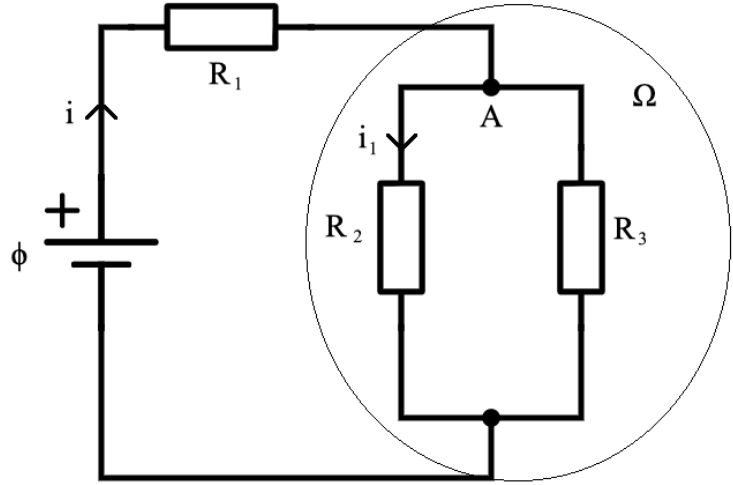

Figure 5. A simple circuit to demonstrate the application of MinEP principle to determine the currents in the subsystem $\Omega$.

The total rate of the entropy production in $\Omega$ is given by

$$T\frac{dS_\Omega}{dt} = i_1^2 R_2 + (i-i_1)^2 R_3 \tag{36}$$

The rate of entropy production is a function of the independent variable $i_1$ and, hence, the critical point can be found by

$$\frac{d}{di_1}\left(T\frac{dS_\Omega}{dt}\right) = 2i_1 R_2 - 2(i-i_1) R_3 = 0 \tag{37}$$

The equation (37) gives us the correct distribution of currents in the subsystem. We shall illustrate the second derivative test to determine if the critical point is actually a minimum and the limitations of applying the MinEP principle in a later section.

5. **Maximum Entropy Production principle**

While the second law of thermodynamics states that the entropy of an isolated system reaches a maximum when it attains its equilibrium state, the Maximum Entropy Production (MaxEP) principle states that the rate of increase of entropy is also a maximum. There have recently been several successful applications of MaxEP principle in diverse areas of science such as physics, chemistry and biology. A review of various applications of MaxEP principle and its historical background can be found in an excellent review article by Martyushev and Seleznev [35]. Paško Županović, Davor Juretić and Srećko Botrić have recently proved that for a linear planar electric network, KVL can be derived from the principle of Maximum Entropy Production (MaxEP) [37, 38]. They consider a linear planar network consisting of active parts such as ideal voltage sources and passive parts such as resistors that convert energy of active parts into Joule heat that is given off to the environment, maintained at a constant temperature $T_a$. We shall now present their idea using the simple circuit shown in figure 6.

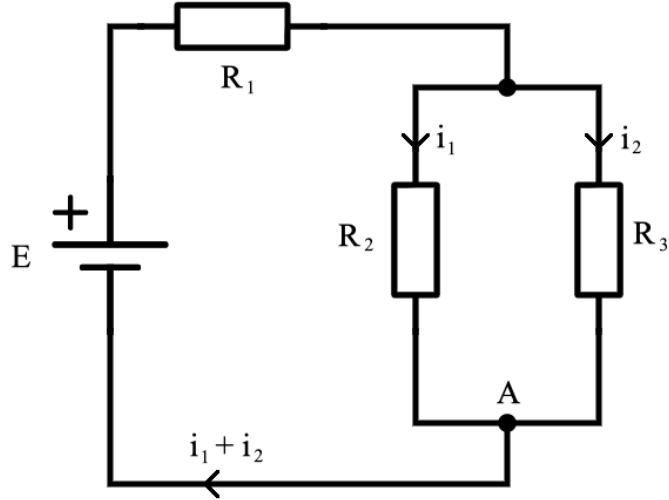

Figure 6. A simple circuit to demonstrate the application of the MaxEP principle to solve for the steady-state distribution of currents.

The validity of KVL in a DC circuit means that the first law of thermodynamics is valid in each loop. We shall, however, assume that the first law of thermodynamics is only valid for the entire circuit. Then the global conservation of energy means that the rate at which energy $W$ is delivered to passive parts of the circuit should equal the rate of dissipation of energy $Q$ into heat in them. The constraint equation then is given by

$$\frac{dW}{dt} - \frac{dQ}{dt} = 0 \tag{38}$$

If the circuit is placed in an environment that is maintained at a constant temperature $T_a$ and a current $i_i$ flows through the passive element $R_i$ of the circuit, the rate of entropy produced in the passive elements of the circuit that is given off to the environment is given from equation (35) as

$$\frac{dS_i}{dt} = \frac{R_i i_i^2}{T_a} \tag{39}$$

The total rate of entropy produced in the passive elements of the circuit is then given by the additive nature of the entropy as

$$\frac{dS}{dt} = \sum_i \frac{dS_i}{dt} = \frac{1}{T_a}\left[R_1(i_1+i_2)^2 + R_2 i_1^2 + R_3 i_2^2\right] \tag{40}$$

The principle of the Maximum Entropy Production (MaxEP) principle states that the currents, while satisfying KCL at every point in the circuit and subject to the law of conservation of energy in the entire circuit, distribute themselves so as to maximize the total rate of entropy production in the circuit. Hence, we now have the following optimizing problem:

$$\left.\begin{array}{l} \text{maximize: } R_1(i_1+i_2)^2 + R_2 i_1^2 + R_3 i_2^2 \\ \text{subject to: } E(i_1+i_2) - R_1(i_1+i_2)^2 - R_2 i_1^2 - R_3 i_2^2 = 0 \end{array}\right\} \quad (41)$$

Solving the optimization problem shown in equations (41) results in the correct distribution of steady-state currents in the circuit. The details of this particular problem and the Lagrange method to solve it are presented in [39] and we shall not go into the further details of the solution in this paper.

## 6. Limitations of the MaxEP and MinEP principles

The MinEP and MaxEP have been used to arrive at a solution for a steady state operation of DC circuits. We shall now take a closer look at the differences between MinEP and MaxEP in solving DC circuits and also describe the nature of these two principles. The MinEP and MaxEP variational principles for describing the dynamics of the open systems are not mutually opposed, but they are just different principles with different constraints. Lucia recently proved that MinEP principle is related to the system while the MaxEP principle is used to describe the interaction of system and its environment [40].

We shall now, using the example of a simple DC circuit, prove the different constraints used in MinEP and MaxEP principles and their limitations. Consider again the circuit shown in figure 6. The correct distribution of the steady-state currents that maximize the rate of total entropy given to the environment are obtained by solving the optimization problem described by equation (41). At steady state, it has been shown that [41]

$$\frac{d^2}{di_1^2}\left(\frac{dS}{dt}\right)\bigg|_{\frac{d\dot{S}}{di_1}=0} = -\frac{2(R_2+R_3)}{T_a} < 0 \quad (42)$$

Analogous relationship can be obtained for the other independent variable $i_2$. Hence, the rate of entropy production is a maximum. Next, we consider the solution provided by the MinEP principle for the circuit shown in figure 5. The solution given by solving linear equation (37) represents a critical point of the function $T_a dS_\Omega / dt$. In order to prove that the critical point is a local minimum, we prove that $D > 0$, where $D$ is defined as

$$D = \frac{d^2}{di_1^2}\left(T\frac{dS_\Omega}{dt}\right) \quad (43)$$

Substituting the value of $T_a dS_\Omega / dt$ from equation (36), we have $D = 2(R_3 + R_2) > 0$. Hence, $dS_\Omega / dt$ has a minimum value at the steady state distribution of the currents in the circuit. We can also modify the circuit shown in figure 5 to determine steady-state currents in the resistors $R_2$ and $R_3$, as demonstrated by T Christen [34]. Since MinEP refers to minimization of entropy production rate of a system at a given input current *I*, we must have a fixed current flowing through the resistors $R_2$ and $R_3$. This can be done by converting the series combination of voltage source $\phi$ and resistor $R_1$ into a constant current source by letting $\phi \to \infty$ and $R_1 \to \infty$ such that $\phi/R_1 = I$. If we denote the entropy production rate in the parallel combination of $R_2$ and $R_3$ (subsystem $\Omega$) by $\dot{S}_\Omega$ and the total rate of entropy transferred to the environment by $\dot{S}$, we note, from the additive nature of entropy, that

$$\frac{d\dot{S}}{di_1} = \frac{d\dot{S}_\Omega}{di_1} + \frac{2R_1 I}{T}\frac{dI}{di_1} \tag{44}$$

Since for, a constant current source *I*, $dI/di_1 = 0$, we then have that

$$\frac{d\dot{S}}{di_1} = \frac{d\dot{S}_\Omega}{di_1} \tag{45}$$

Hence, the MinEP principle does not take entropy generated by the resistor $R_1$ into the account and this criticism to the application of the MinEP principle in electric circuits was raised by T Chirsten in his paper [34]. He also showed that MaxEP principle can be used to obtain correct distribution of steady-state currents for systems far from equilibrium and when non-linear systems are present. The entropy production principles cannot be used when the temperatures of the resistors and generator are different. Stijn Bruers *et al* showed recently using dynamic fluctuation theory that MinEP and MaxEP principles cannot be applied to temperature-inhomogeneous circuits [35].

## 7. Conclusions and implications for teaching

The DC circuits are often introduced in electrical engineering or basic physics courses and Kirchhoff's laws are stated, without a general proof, as a useful technique to find their distribution at the steady-state operation of the circuit. The KVL requires the understanding of the concept of traversal in a (electrical) loop and it also requires the students to memorize the different signs of potential difference to be assigned for sources and resistors. While KCL can be understood easily by students, they often have difficulty applying KVL due to its abstract nature. We believe that the variational principles can offer a useful alternative teaching methods solving DC circuits without the use of KVL.

The variational principles in physics, such as the principle of least action, are often introduced in advanced mechanics courses. Recently, many authors have argued that least action principle provides a powerful technique for unifying Newton's mechanics, relativity and quantum mechanics and they proposed the advantages of teaching them in the introductory courses [42-49]. Building on their work, Thomas A. Moore presented changes that can be implemented in the upper-level physics curriculum to teach Lagrangian methods and variational methods that are also very useful in contemporary research [50]. The variational methods in DC circuits, as demonstrated in this review article, can be another useful opportunity to teach students the techniques of variational principles in a way that can appeal to the physical understanding of the students.

Finally, DC circuits offer a unique opportunity to introduce the concepts and techniques of irreversible thermodynamics. We believe that introduction of MaxEP and MinEP principles and the concepts of fluxes and forces in irreversible thermodynamics through DC circuits to students who are familiar with solving DC circuits can be advantageous. While we have not provided a detailed curriculum of upper-level courses for incorporating the teaching of variational principles in DC circuits, we believe that such a curriculum is possible and can be very useful.

**Acknowledgements**

I wish to thank the two anonymous referees for their comments and insights that helped in improving the quality of the manuscript. I would also like to thank Ms. Tanya Purwar, Mr. Sameep Shah and Ms. Vasudha Kapre for their helpful discussions.